\documentclass[11pt]{article}
 \usepackage{commands}

\title{Efficient Decoding of Double-circulant and Wozencraft Codes \\ from Square-root Errors}

\author{Oren Dubin, Noam Oz, and Noga Ron-Zewi \thanks{Emails: \texttt{oren.dubin@gmail.com, noamoz296@gmail.com, nronzewi@ds.haifa.ac.il}. Research supported by  ISF grant 735/20, and by the European Union (ERC, ECCC, 101076663). Views and opinions expressed are however those of the author(s) only and do not necessarily reflect those of the European Union or the European Research Council. Neither the European Union nor the granting authority can be held responsible for them.}}

\date{University of Haifa}

\begin{document}

\maketitle

\begin{abstract}

We present efficient decoding algorithms from square-root errors for two known families of double-circulant codes: A construction based on Sidon sets (Bhargava, Taveres, and Shiva, \emph{IEEE IT 74}; Calderbank, \emph{IEEE IT 83}; Guruswami and Li, \emph{IEEE IT 2025}),
and a construction based on cyclic codes (Chen, Peterson, and Weldon, \emph{Information and Control 1969}).
We further observe that the work of Guruswami and Li implicitly gives a transformation from double-circulant codes of certain block lengths to Wozencraft codes which preserves that distance of the codes, and we show that this transformation also preserves efficiency of decoding. By instantiating this transformation with the first family of double-circulant codes based on Sidon sets, we obtain an explicit construction of a Wozencraft code that is efficiently decodable from square-root errors. We also discuss limitations on instantiating this transformation with the second family of double-circulant codes based on cyclic codes. 

\end{abstract}

\section{Introduction}

In this work, we present efficient decoding algorithms from square-root errors for two known families of double-circulant codes, and use these to provide an explicit construction of a Wozencraft code that can be efficiently decoded from square-root errors. Next we elaborate on each of these results.

\paragraph{Double-circulant codes.}
A $k \times k$ matrix $A$ over a field $\F$ is called \emph{circulant} if each column is obtained by taking a cyclic shift of the previous column, that is, it is of the form:
  $$  A=\begin{pmatrix}
        a_0 & a_{k-1}  & \cdots & a_{1}\\
        a_{1} & a_0  & \cdots & a_{2}\\
        \vdots & \vdots  & \ddots & \vdots\\
        a_{k-1} & a_{k-2}  & \cdots & a_0 
    \end{pmatrix}.
    $$
A \emph{double-circulant code} (also known as a \emph{quasi-cyclic code}) is a linear code $D \subseteq \F^{2k}$ over some finite field $\F$ which has a generator matrix of the form  $G=\begin{pmatrix}
        I_k \\ A 
    \end{pmatrix},$
    where $I_k$ is the $k \times k$ identity matrix and $A$ is a $k \times k$ circulant matrix 
\cite[Page 497]{MS77}. 

Several works in the past investigated the properties and constructions of double-circulant codes  \cite{CPW69, BTS74, Kasami74, Calder83, RL90} (see also \cite[Chapter 16, Section 7]{MS77}). 
In particular, it was shown in \cite{CPW69, Kasami74} that random double-circulant codes achieve the Gilbert-Varshamov bound with high probability. With regard to explicit constructions, Bhargava, Tavers, and Shiva \cite{BTS74} suggested a construction based on \emph{Sidon difference sets} \cite{Sidon32}, and in a follow-up work  \cite{Calder83}, Calderbank  showed that this construction gives an asymptotic family of double-circulant codes of distance $\Omega(\sqrt{k})$. Roth and Lempel \cite{RL90} presented another explicit construction of an asymptotic family of double-circulant codes of square-root distance, based on the Fourier Transform, and also devised an efficient algorithm for decoding these codes from square-root errors. 

Chen, Peterson, and Weldon \cite{CPW69} suggested a general method for constructing double-circulant codes from cyclic codes, where the distance of the resulting double-circulant codes is lower bounded by the minimum of the distance and the dual distance of the  underlying cyclic codes. Chen, Peterson, and Weldon did not discuss how to instantiate their transformation to obtain an asymptotic family of codes of good distance, 
but we observe that instantiating their transformation with either BCH codes \cite{BR60} or punctured Reed-Muller codes \cite{Muller54, Reed54, KLP68} also leads to an asymptotic family of double-circulant code with square-root distance. 

As far as we are aware, there are no known explicit constructions of an asymptotic family of double-circulant codes of distance  $\omega(\sqrt{k})$. Also, to the best of our knowledge, prior to our work the only asymptotic family of double-circulant codes that was known to be  efficiently decodable from square-root errors was the Fourier-based codes of \cite{RL90}. In this work, we devise efficient decoding algorithms from square-root errors for the two other families of double-circulant codes mentioned above \cite{BTS74, CPW69}.  Next we describe our methods for decoding these codes.

\paragraph{Decoding double-circulant codes.}
To decode the double-circulant codes based on Sidon difference sets of \cite{BTS74}, we show in Section  \ref{sec:Sidon_to_DC} a general result that applies to any linear code $C \subseteq \F^{2k}$ with a \emph{systematic} generator matrix of the form 
 $G=\begin{pmatrix}
        I_k \\ A 
    \end{pmatrix},$
    where $I_k$ is the $k \times k$ identity matrix and $A$ is a $k \times k$ \emph{design matrix} (that is not necessarily circulant). Specifically, we say that $A$ is a $(d,b)$-\emph{design matrix} if any column of $A$ has weight at least $d$, and the intersection of the supports of any pair of columns of $A$ has size at most $b$. We show that if $A$ is a $(d,b)$-design matrix, then $C$ has distance at least $\frac d b$, and it can be efficiently decoded from $\frac{d} {2b}$ errors. We then show that the matrix $A$ which defines the Sidon-based codes of  \cite{BTS74} is an $(\Omega(\sqrt{k}), 2)$-design matrix, and consequently these codes can be efficiently decoded from $\Omega(\sqrt{k})$ errors. This also gives an alternative proof for the square-root distance of these codes that was first shown in \cite{Calder83}. 
    
    To decode the double-circulant codes based on cyclic codes of \cite{CPW69}, we show in Section \ref{sec:cyc_to_circ} that if both the underlying cyclic codes and their duals are efficiently decodable from $e, e^\perp$ errors, respectively, then the resulting double-circulant code is efficiently decodable from $\min\{e,e^\perp\}$ errors. We then show that instantiating this construction with the punctured Reed-Muller codes, and using Reed's majority logic decoder \cite{Reed54}, this leads to an asymptotic family of double-circulant codes that are efficiently decodable from square-root errors.\footnote{
    Instantiating the construction with the BCH codes will not immediately lead to an efficient decoding algorithm since dual BCH codes are not known to be efficiently decodable. }

\paragraph{Wozencraft codes.}
Our motivation for studying double-circulant codes comes from the recent work of Guruswami and Li \cite{GL25} that presented an explicit construction of a \emph{Wozencraft code} of square-root distance, based on Sidon sets. A \emph{Wozencraft code} is a linear code of the form $$W_\alpha = \left\{ (m,\alpha\cdot m) \mid m \in \F_{q^k} \right\} \subseteq (\F_q)^{2k}$$ for some $\alpha \in \F_{q^k}$, where one interprets elements of $\F_{q^k}$ as strings in $(\F_q)^k$ via some fixed $\F_q$-linear bijection between these vector spaces. It is well-known that for a random $\alpha \in \F_{q^k}$, the Wozencraft code $W_\alpha$ achieves the Gilbert-Varshamov bound with high probability, and this property was used by Justesen \cite{Just72} to obtain the first strongly explicit asymptotically good binary codes, by concatenating an outer Reed-Solomon code with inner Wozencraft codes for different values of $\alpha$. More recently, Wozencraft codes have also found other applications to the construction of covering codes \cite{PZ20} and write-once-memory codes \cite{Shpilka13, Shpilka14}. 

An interesting question is to explicitly find some $\alpha \in \F_{q^k}$ for which the Wozencraft code $W_\alpha$ has good distance, ideally the Gilbert-Varshamov bound, or more modestly, distance $\Omega(k)$. While we are still far from achieving either goals, very recently Guruswami and Li \cite{GL25} have shown how to explicitly find some $\alpha \in \F_{q^k}$ for which the Wozencraft code $W_\alpha$ has distance $\Omega(\sqrt{k})$. Their construction was based on Sidon sets, and we observe that their construction can be abstracted as follows. First, we observe  that their methods implicitly give a transformation from double-circulant codes of certain block lengths to Wozencraft codes which preserves the distance of the codes. The construction of \cite{GL25} can then be obtained 
by instantiating this transformation with the Sidon-based double-circulant codes of \cite{BTS74,Calder83}.

In Section \ref{subsec:circ_to_wozen}, we describe the transformation from double-circulant to Wozencraft codes that is implicit in \cite{GL25}, and we further show that this transformation also preserves efficiency of decoding. We then show in Section \ref{subsec:sidon_to_wozen}
that
instantiating this transformation with the sidon-based double-circulant codes of  \cite{BTS74}, and using our new decoding algorithm from square-root errors for these codes, 
gives an explicit construction of a Wozencraft code that can be efficiently decoded from square-root errors. 
Unfortunately, the transformation from double-circulant to Wozencraft codes requires that the double-circulant codes have a certain block-length, and
this requirement prevents us from instantiating this transformation with the construction of double-circulant codes based on cyclic codes of \cite{CPW69}, as well as the 
Fourier-based construction of double-circulant codes of \cite{RL90}. We discuss these limitations and possible approaches to overcome them in Section \ref{subsec:cyc_to_circ_limit}.

\paragraph{Open questions.}

An interesting open question raised by this work, as well as prior work \cite{GL25}, is to  explicitly construct a Wozencraft code of distance $\omega(\sqrt{k})$. Our work highlights that possible approaches towards this goal may be to explicitly construct a double-circulant code of distance  $\omega(\sqrt{k})$, or cyclic codes  so that both their distance and dual distance is $\omega(\sqrt{k})$, of appropriate block length. We are not aware of any such constructions (even without the restriction on the block length). 

In fact, while for double-circulant codes we know that random constructions achieve the Gilbert-Varshamov bound with high probability (and in particular, have distance $\Omega(k)$) \cite{CPW69, Kasami74}, we are not aware of any cyclic codes with both distance and dual distance $\omega(\sqrt{k})$. Moreover, a well-known conjecture postulates that there do not exist asymptotically-good cyclic codes, and in Proposition \ref{prop:cyclic_conj}, we show that this conjecture implies that there do not exist cylic codes so that both their distance and dual distance is $\Omega(k)$. We thus would like to put forward the following conjecture, which can be viewed as an intermediate step towards the resolution of the conjecture about the non-existence of asymptotically-good cyclic codes.

\begin{conjecture}
For any fixed prime power $q$, there does not exist an infinite family of cyclic codes over $\F_q$ with both constant relative distance and constant relative dual distance.
\end{conjecture}

\paragraph{Paper organization.} 
In Section \ref{sec:prelim}, we begin with setting some notation, and providing the formal definitions of the codes we study.
In Sections \ref{sec:Sidon_to_DC}, we present an efficient decoding algorithm for the Sidon-based double-circulant codes of \cite{BTS74}, and in Section \ref{sec:cyc_to_circ}, we present an efficient decoding algorithm for
the double-circulant codes based on cylic codes of \cite{CPW69}. In Section \ref{subsec:circ_to_wozen}, we describe the transformation from double-circulant to Wozencraft codes that is implicit in \cite{GL25}, and show that this transformation also preserves efficiency of decoding. Then in Section \ref{subsec:sidon_to_wozen}, we show how to instantiate this transformation with the Sidon-based double-circulant codes of \cite{BTS74}, together with our efficient decoding algorithm for these codes, to obtain 
an explicit construction of a Wozencraft code that can be efficiently decoded from square-root errors. Finally, in Section \ref{subsec:cyc_to_circ_limit}, we discuss limitations on instantiating this transformation with the double-circulant codes based on cyclic codes of \cite{CPW69}, as well as the Fourier-based double-circulant codes of \cite{RL90}.

\section{Preliminaries}\label{sec:prelim}

Let $\Sigma$ be a finite alphabet. For strings $u,v \in \Sigma^n$, we define their \textsf{Hamming distance} $\Delta(u,v)$ as the number of indices $i \in [n]$ so that $u_i \neq v_i$. Let $\F$ be a finite field. For a string $w \in \F^n$, we define its \textsf{Hamming weight} $\wt(w)$  as the number of indices $i \in [n]$ so that $w_i \neq 0$. 
We let $\F_{\leq k}[x]$ ($\F_{<k}[x]$, respectively) denote the collection of polynomials of degree at most (smaller than, respecitvely) $k$ over $\F$. 
Given a string $w=(w_0, \ldots, w_{k-1}) \in \F^k$, we let $w(x)$ denote the polynomial in $\F_{<k}[x]$ whose coefficients are the entries in $w$, that is, $w(x):=\sum_{i=0}^{k-1} w_i x^i \in \F_{<k}[x]$. Conversely, given a polynomial $f(x) \in \F_{<k}[x]$, we let $f \in \F^k$ denote its length $k$ coefficient vector. 

For a prime power $q$, we denote by $\F_q$ the finite field of $q$ elements, and we denote by $\F_q^* = \F_q \setminus \{0\}$ its multiplicative group. 
Let $p$ be a prime, and let $n$ be an integer. We say that $n$ is a 
 \textsf{primitive root modulo $p$} if $n \;(\bmod \; p)$ generates $\F_p^*$, that is, it 
 has multiplicative order $p-1$ in $\F_p^*$. 

\begin{proposition}[\cite{GL25}, Proposition 2.4]\label{prop:primitive}
Let $q$ be a prime power, and let $k$ be a prime so that $q$ is a primitive root modulo $k$.  Then the polynomial $p_k(x):=\sum_{i=0}^{k-1}x^i$ is an irreducible polynomial over $\F_q$. 
\end{proposition}

\subsection{Error-correcting codes}

An \textsf{(Error-correcting) code} is a subset $C\subseteq\Sigma^{n}$.  We call $\Sigma$ and $n$ the \textsf{alphabet} and the
\textsf{block length} of the code, respectively, and the elements of $C$ are called \textsf{codewords}. 
 The \textsf{rate}
of a code $C \subseteq \Sigma^n$ is the ratio $R(C):=\frac{\log|C|}{n \cdot \log |\Sigma|}$. The \textsf{(Hamming) distance} of $C$ is $\Delta(C):= \min_{c \neq c' \in C} \Delta(c,c')$ 
and its \textsf{relative distance} is $\delta(C):= \frac{\Delta(C) } {n}$.  

We say that a code $C\subseteq\Sigma^{n}$ is \textsf{linear} if $\Sigma = \F$ 
for some finite field $\F$, and $C$ is a linear subspace of $\F^n$. 
For a linear code $C\subseteq\F^n$, we have that $R(C)=\frac{\dim(C)}{n}$ and 
$\Delta(C)=\min_{0 \neq c \in C} \wt(c)$. 
A \textsf{generator matrix} for a linear code $C \subseteq \F^n$ of dimension $k$ is a (full-rank) matrix $G \in \F^{n \times k}$ so that $\img(G) = C$, and a \textsf{parity-check matrix} for $C$ is a (full-rank) matrix $H \in \F^{ (n-k) \times n}$ so that $\ker(H)= C$. 
The \textsf{dual code} of $C$ is the code $C^\perp \subseteq \F^n$ containing all strings $c' \in \F^{n}$ satisfying that $\sum_{i=1}^n c'_i \cdot c_i = 0$ for all $c \in C$. 
It follows by definition that $(C^{\perp})^{\perp} = C$, and that $H$ is a parity-check matrix for $C$ if and only if $H^T$ is a generator matrix for $C^{\perp}$. 

Next we formally define the codes we study.

\paragraph{Cyclic codes.}
We say that a linear code $C\subseteq \F^n$ is \textsf{cyclic} if for every codeword $(c_0, c_1, \dots, c_{n-1})\in C$, it holds that $(c_{n-1},c_0,c_1\dots, c_{n-2})\in C$. 
It is well-known that for a cyclic code $C \subseteq \F^n$ of dimension $k$, 
 there exists a unique monic \textsf{generator polynomial} $g(x) \in \F_{\leq n-k}[x]$  so that the codewords of $C$ are all length $n$ coefficients vectors of polynomials of the form  
 $g(x)\cdot f(x) \in \F_{<n}[x]$ for $f(x) \in \F_{<k}[x]$. Moreover, $g(x)$ divides $x^n-1$. 
The \textsf{check polynomial} $h(x) \in \F_{\leq k}[x]$ is defined as the polynomial which satisfies that $g(x)\cdot h(x)=x^n-1$. 

For a vector $w \in \F^n$, let $w^{\rev} \in \F^n$ denote the reverse vector of $w$, and for a subset $C \subseteq \F^n$, let $C^{\rev} =\{c^{\rev} \mid c \in C\}$. Note that if $C$ is a cyclic code of distance $d$, then so is $C^{\rev}$. 
For a polynomial $h(x) \in \F_{\leq k}[x]$, let $h^{\rev}(x)\in \F_{\leq k}[x]$ denote the polynomial whose coefficients appear in reverse order, that is, $h^{\rev}(x) =x^k \cdot h(\frac 1 x)$.

\begin{fact}\label{fact:cyclic_dual}
Let $C \subseteq \F^n$ be a cyclic code of dimension $k$ with check polynomial $h(x) \in \F_{\leq k}[x]$. Then $C^\perp$ is a cyclic code with generator polynomial  $h^{\rev}(x)$.
\end{fact}

One family of cyclic codes that will be useful for us are punctured Reed-Muller codes. 
For positive integers $r<m$, the \textsf{Reed-Muller code}  $\RM(r,m)$ is a binary linear code of block
length $2^m$ which associates with any  polynomial $f(x_1, \ldots, x_m) \in (\F_2)_{\leq r}[x_1, \ldots, x_m]$ a codeword $(f(a))_{a \in \F_2^m}$. 

\begin{fact}[\cite{Muller54, Reed54}]\label{fact:rm}
The Reed-Muller code $\RM(r,m)$ has distance $2^{m-r}$, and is  decodable up to half the minimum distance in time $\poly(2^m)$. Moreover, $\RM(r,m)^\perp = \RM(m-r-1,m)$. 
 \end{fact}

The \textsf{punctured RM code} $\RM^*(r,m)$ is the code obtained from $\RM(r,m)$ by omitting the entry which corresponds to $a=0$. It is known that $\RM^*(r,m)$ is cyclic for a certain ordering of the codeword entries.

\begin{proposition}[\cite{KLP68}]\label{prop:prm}
Let $m,r$ be positive integers, let
$\varphi: \F_{2^m} \to \F_2^m$ be an $\F_2$-linear bijection, and let $\alpha$ be a generator of the multiplicative group of the field $\F_{2^m}$. Then the punctured RM code 
$\RM^*(r,m)$ is cyclic when the  evaluation points are ordered as $\varphi(\alpha^i)$ for $i=0,1, \ldots, 2^m-1$.
\end{proposition}

\paragraph{Circulant codes.}
 A \textsf{Circulant matrix} over a field $\F$ is a square matrix over $\F$ where each column is obtained by taking a cyclic shift of the previous column, that is, it is of the form:

$$
    A=\begin{pmatrix}
        a_0 & a_{k-1}  & \cdots & a_{1}\\
        a_{1} & a_0  & \cdots & a_{2}\\
        \vdots & \vdots  & \ddots & \vdots\\
        a_{k-1} & a_{k-2}  & \cdots & a_0 
    \end{pmatrix}.
$$

\begin{claim}\label{clm:circulant}
Let $a := (a_0, a_1, \ldots, a_{k-1}) \in \F^k$ denote the first column of $A$. Then for any $m \in \F^k$, $A \cdot m$ is the length $k$ coefficient vector of the polynomial
$a(x) \cdot m(x) \; (\bmod \; x^k-1) \in \F_{<k}[x]$.
\end{claim}

We say that a linear code $D \subseteq \F^{t\cdot k}$ of dimension $k$ is $t$-\textsf{circulant} if it has a
 generator matrix of the form
 $$G=\begin{pmatrix}
        I_k \\ A_1 \\ \vdots \\ A_{t-1}
    \end{pmatrix},$$
where $I_k$ is the $k \times k$ identity matrix, and $A_1, \ldots, A_{t-1}$ are $k \times k$ circulant matrices over $\F$. 
In the special case that $t=2$, we say that $D$ is \textsf{double circulant}.

\paragraph{The Weldon and Wozencraft ensembles \cite{Massey63, Weldon73}.}
Let $q$ be a prime power, and let $k \geq 1$ be a positive integer. We say that  $W \subseteq \F_q^{t \cdot k}$ is a $t$-\textsf{Weldon code} if there exist $\alpha_1, \ldots, \alpha_{t-1} \in \F_{q^k}$ and an $\F_q$-linear bijection $\varphi: \F_{q^k} \to \F_q^k$ so that 
$$W=\left\{ (\varphi(m), \varphi(\alpha_1 \cdot m), \ldots, \varphi(\alpha_{t-1} \cdot m)) \mid m \in \F_{q^k}\right\}.$$
In the special case that $t=2$, we say that $W$ is a \textsf{Wozencraft code}.

\section{Decoding double-circulant codes based on Sidon sets}\label{sec:Sidon_to_DC}

In this section, we present an efficient decoding algorithm from square-root errors for the double-circulant codes based on Sidon sets of \cite{BTS74}. This also gives an alternative proof for the square-root bound on the distance of these codes that was first proved in \cite{Calder83}, and also recently implicitly shown in \cite{GL25}. 
To formally state the result, we first provide the formal definition of a Sidon set. 

\begin{definition}[Sidon set,  \cite{Sidon32}]\label{def:sidon}
Let $S=\{s_1,s_2,\dots ,s_d\} \subseteq \N$  be a subset of $d$ distinct integers.
We say that $S$ is a  \textsf{Sidon set} if 
its pairwise differences are all distinct, that is, for any $i,j,k,\ell\in [d]$ so that $i\neq j$ and $k\neq \ell$, it holds that $s_i-s_j=s_k-s_\ell$ if and only if $i=k$ and $j=\ell$.
\end{definition}

For the transformation from double-circulant to Wozencraft codes, presented in Section \ref{sec:circ_to_wozen}, we shall need a slightly stronger property out of the double-circulant codes, namely that they have a high \emph{balanced weight}, as per the following definition. 

\begin{definition}[Balanced weight]\label{def:balanced_weight}
Let $\F$ be a finite field. 
We define the \textsf{balanced weight} $\wt_\bal(w)$ of a string $w \in \F^n$ as $\min_{\alpha \in \F} |\{i \in [n] \mid w_i \neq \alpha\}|$. Note that $\wt_\bal(w) \geq d$ if any element $\alpha \in \F$ appears at most $n-d$ times in $w$, and that $\wt(w) \geq \wt_\bal(w)$. 
We say that a code $C \subseteq \F^k$ is $d$-\textsf{balanced} if $\wt_\bal(c) \geq d$ for any non-zero codeword $c$. 
For an integer $t \geq 2$, we say  that a code $C \subseteq \F^{t \cdot k}$ is $(t,d)$-\textsf{balanced} if for any non-zero codeword $c=(c_0, c_1, \ldots,c_{t-1} ) \in C$, where $c_0, c_1, \ldots, c_{t-1} \in \F^k$, we have that $\wt(c_0)+\sum_{i=1}^{t-1}\wt_\bal(c_i) \geq d$. 
\end{definition}

Our main result in this section is the following.

\begin{theorem}[Efficient decoding of double-circulant codes based on Sidon sets]\label{thm:Sidon_to_DC}
Suppose that there exists an explicit construction of an infinite sequence of sets $\mathcal{S}=\{S_k\}_{k \in I}$, where $S_k \subseteq [k]$ is a Sidon set of size $d=d(k)$.
Then for any prime power $q$, there exists an explicit family of codes $\mathcal{D}=\{D_k\}_{k\in I}$, where $D_k \subseteq \F_q^{2 k}$ is a  double-circulant code of distance at least $\frac d 2+1$ that can be decoded from less than $\frac d 4$ errors in time $\poly(k, \log q)$.
Moreover, $D_k$
is $(2,d')$-balanced for $d'=\min\{\frac {d} {2}+1, \frac{k} {d}\}$. 
\end{theorem}

We prove the above theorem in Section \ref{subsec:sidon_to_DC} below.
This theorem can be instantiated with the following construction of Sidon sets from \cite{BC60}.

\begin{theorem}[Explicit Sidon Set, \cite{BC60}]\label{thm:sidon}
For any prime power $q$, there exists a Sidon set $S \subseteq [q^2]$  of size $q$ that can be constructed in time $\poly(q)$.
\end{theorem}

In particular, for any $k \in \N$, one can apply the above theorem with a  prime $q$ so that $\frac 1 2 \sqrt{k} \leq q \leq \sqrt{k}$ (such a prime exists by the Bertrand-Chebyshev Theorem \cite{Cheb1850}), and obtain a Sidon set $S \subseteq [k]$ of size $\Theta(\sqrt{k})$ that can be constructed in time $\poly(k)$. Theorem \ref{thm:Sidon_to_DC} then implies the following corollary.

\begin{corollary}\label{cor:DC}
For any prime power $q$, 
there exists an explicit family of codes $\mathcal{D}=\{D_k\}_{k\in \N}$, where 
$D_k \subseteq \F_q^{2 k}$ is a  double-circulant code of distance at least $d=\Omega(\sqrt{k})$ that can be decoded from less than $\frac d 2$ errors in time $\poly(k, \log q)$. Moreover, $D_k$ is $(2,\Omega(\sqrt{k}))$-balanced. 
\end{corollary}

\subsection{Proof of Theorem \ref{thm:Sidon_to_DC}}\label{subsec:sidon_to_DC}

To prove Theorem \ref{thm:Sidon_to_DC}, we first present the construction of double-circulant codes from Sidon sets, then analyze their distance, and finally present a decoding algorithm for these codes. 

\paragraph{Construction.}
Fix $k \in I$, the code $D_k$ is constructed as follows. Let $a=(a_1, a_2, \ldots, a_k) \in \{0,1\}^k$ be the indicator vector of $S_k \subseteq [k]$, that is, $a_i=1$ if and only if $i \in S_k$. Let $A$ be the binary $k \times k$ circulant matrix whose first column is $a$, and let $D_k$ be a double-circulant code with generator matrix  $G=\begin{pmatrix}
        I_k \\ A
    \end{pmatrix}$.

\paragraph{Distance and balanced weight.} 

Our distance proof relies on the following notion of a \emph{design matrix}.

\begin{definition}[Design matrix]
We say that a matrix $A$ over a field $\F$ is a
$(d,b)$-\textsf{design matrix} if each column of $A$ has weight at least $d$, and the intersection of the supports of any pair of columns of $A$ has size at most $b$. 
\end{definition}

\begin{lemma}\label{lem:sidon_1}
Suppose that $C \subseteq \F_q^{2k}$ is a linear code with generator matrix of the form $G=\begin{pmatrix}
        I_k \\ A
    \end{pmatrix}$, where $A$ is a $k \times k$ matrix over $\F_q$. Then if $A$ is a $(d,b)$-design matrix, then $C$ has distance at least $\frac{d} {b}+1$. Moreover, if $A$ is circulant, and each row in $A$ has weight at most $r$, then $A$ is $(2,d')$-balanced for $d'=\min\{\frac {d} {b}+1, \frac{k} {r}\}$. 
\end{lemma}

\proof
Let $c=(c_0,c_1) \in C$ be a non-zero codeword, where $c_0, c_1 \in \F_q^k$, we shall show that $wt(c) \geq \frac{d} {b} +1$. To see this, note that $c$ is a non-zero linear combination of columns of $G$. Suppose that this linear combination has $\ell$ non-zero entries $i_1, \ldots, i_\ell \in [k]$ for some $\ell \in [k]$.

Then by the structure of $G$, we have that $\wt(c_0) = \ell$. Moreover, since $A$ is a $(d,b)$-design matrix, we have that the $i_1$ column of $G$ has weight at least $d$, and the support of any other column from $i_2, \ldots, i_\ell$ intersects the support of $i_1$ in at most $b$ entries. Consequently, we have that $\wt(c_1) \geq \max\{0,d - (\ell-1) \cdot b\}$. But this implies in turn that 
$$\wt(c) \geq \ell+ \max\{0,d - (\ell-1) \cdot b\} =\max\{\ell,d +b - \ell(b-1)\} .$$
Finally, note that the righthand expression is minimized when $\ell = \frac{d} {b}+1$, in which case it equals $\frac{d} {b}+1$. 

For the moreover part, 
let $a \in \F_q^k$ denote the first column of $A$, and note that since $A$ is circulant, $\wt(a) \leq r$. Let $c=(c_0,c_1) \in C$ be a non-zero codeword, where $c_0, c_1 \in \F_q^k$. 
Then by Claim \ref{clm:circulant} and the structure of $G$, we have that $c_1$ is the length $k$ coefficient vector of $a(x) \cdot c_0(x) \; (\bmod\; x^k -1)$. Suppose that $\wt(c_0)=\ell$. Then 
$\wt(c_1) \leq \wt(a) \cdot \wt(c_0) \leq r \cdot \ell$, and so $c$ has at least $\max\{0,k-\ell \cdot r\}$ zero entries. But this means that 
$$\wt(c_0) + (k- \wt(c_1)) \geq \ell +\max\{0,k-\ell \cdot r\} = \max\{\ell,k-\ell \cdot (r-1)\}, $$
where the righthand expression is minimized when $\ell = \frac kr$, in which case it equals $\frac kr$. Finally, note that
$$\wt_\bal(c) \geq \min \left\{\wt(c_0) + \wt(c_1), \wt(c_0) + (k- \wt(c_1)) 
\right\} \geq \min \left\{\frac{d} {b} +1, \frac k r\right\}.$$

\begin{flushright} $\blacksquare$ \end{flushright}

The next lemma says that the circulant matrix $A$ which defines the code $D_k$ is a $(d,2)$-design matrix.

\begin{lemma}\label{lem:sidon_2}
Let $a=(a_1, a_2, \ldots, a_k) \in \{0,1\}^k$ be the indicator vector of a Sidon set $S \subseteq [k]$ of size $d$, and let $A$ be the binary $k \times k$ circulant matrix whose first column is $a$. Then $A$ is a $(d,2)$-design matrix. 
\end{lemma}

\proof

Since $A$ is circulant and $\wt(a)=d$, we clearly have that each column of $A$ has weight $d$. Next we show that the intersection of the supports of any pair of columns of $A$ has size at most $2$.

Fix $1\leq i<j \leq k$, we shall show that the intersection of the supports of the 
 $i$-th and $j$-th columns of $A$ has size at most $2$. 
Suppose on the contrary that  the intersection has size at least $3$. 
Then by the circulant structure of $A$, there exist distinct $s_1,  s_2, s_3 \in S$ and distinct $s'_1, s'_2, s'_3 \in S$ so that
\begin{equation}
    s_1+i\equiv s'_1+j, \;\;s_2+i \equiv s'_2+j, \;\;\text{and}\;\; s_3+i\equiv s'_3+j\; (\bmod \; k),
\end{equation}
or equivalently,
$$
    s_1-s'_1\equiv s_2-s'_2\equiv s_3-s'_3\equiv j-i\;(\bmod \;k).
$$
Furthermore, since $S \subseteq [k]$, we have that $-k+1\leq s_1-s'_1,s_2-s'_2,s_3-s'_3\leq k-1$, therefore they are all equal $j-i$ or $j-i-k$, since $1\leq j-i\leq k-1$. Furthermore, we have that $s_1 \neq s'_1$, $s_2 \neq s'_2$, and $s_3 \neq s'_3$.

Consequently, by the pigeonhole principle, there exist two values out of the three values $s_1-s'_1,s_2-s'_2,s_3-s'_3$ which are the same, without loss of generality, assume that  $s_1-s'_1=s_2-s'_2$. But by the definition of a Sidon set, this implies in turn that $s_1=s_2$, which contradicts the assumption that $s_1, s_2, s_3$ are all distinct.

\begin{flushright} $\blacksquare$ \end{flushright}

Finally, by the definition of $D_k$ and Lemmas \ref{lem:sidon_1} and \ref{lem:sidon_2} above we conclude that $D_k$ has distance at least $\frac d 2+1$, and that it is $(2,d')$-balanced for 
$d'=\min\{\frac {d} {2}+1, \frac{k} {d}\}$.

\paragraph{Decoding.}

Similarly to the distance proof, the decoding algorithm also relies on the notion of matrix design. Specifically, we have the following lemma. 

\begin{lemma}\label{lem:sidon_3}
Suppose that $C \subseteq \F_q^{2k}$ is a linear code with a generator matrix of the form $G=\begin{pmatrix}
        I_k \\ A
    \end{pmatrix}$, where $A$ is a $k \times k$ matrix over $\F_q$. Then if $A$ is a $(d,b)$-design matrix, then $C$  can be decoded from less than $\frac{d} {2b}$ errors in time $poly(k, \log q)$.
\end{lemma}

\proof 

The decoding algorithm  for the code  $C$ is given in Figure \ref{fig:matrix_design_dec} below. For $i \in [k]$, let $S_i \subseteq [k]$ denote the support of the $i$-th column in $A$.

\begin{figure}[h]
  \begin{boxedminipage}{\textwidth} \small \medskip \noindent
    $\;$

    \underline{\textbf{Decoding algorithm  for the matrix design code $C$:}}
\begin{itemize}
\item \textbf{INPUT:} A received word $w=(w_0, w_1) \in \F_q^{2k}$, where $w_0, w_1 \in \F_q^k$ 
\item \textbf{OUTPUT:}  A codeword $c \in C$ so that $\dist(w,c)< \frac{d} {2b}$, or $\bot$ if such a codeword does not exist

\end{itemize}

\begin{enumerate}
\item \label{step:matrix_design_syndrome} Let $y:=A\cdot w_0 -w_1 \in \F_q^k$.
\item \label{step:matrix_design_error}
Compute $z =(z_1, \ldots, z_k) \in \F_q^k$, where for $i =1, \ldots, k$,
$z_i = \maj\{ y_j \mid j \in \supp(S_i)\}$.
\item  \label{step:matrix_design_codeword}
Let $c_0 = w_0 - z \in \F_q^k$ and $c=G \cdot c_0$. 
\item \label{step:matrix_design_output}
 If $\dist(w,c) < \frac{d} {2b}$ return $c$, otherwise return $\bot$. 
\end{enumerate}

  \medskip

  \end{boxedminipage}

\caption{Decoding algorithm for matrix design codes}
\label{fig:matrix_design_dec}
\end{figure}

The running time of the algorithm is clearly $\poly(k, \log q)$. To show correctness, let $w=(w_0, w_1) \in \F_q^{2k}$ be a received word, where $w_0, w_1 \in \F_q^k$, and let 
$c=(c_0, c_1) \in C$ be a codeword so that $\dist(c,w) < \frac{d} {2b}$, where $c_0, c_1 \in \F_q^k$. We shall show that the algorithm will output $c$.
To this end, it suffices to show that the string $z$ which the algorithm computes on Step \ref{step:matrix_design_error}  satisfies that $z = w_0-c_0$, as in this case the algorithm will clearly compute the correct codeword $c$ on Step \ref{step:matrix_design_codeword}.

Let $e=(e_0,e_1):=w-c \in \F_q^{2k}$, where $e_0, e_1 \in \F_q^k$, and note that $\wt(e) < \frac{d} {2b}$ by assumption that $\dist(c,w) < \frac{d} {2b}$. Moreover, we have that
$$y = A\cdot w_0 - w_1 = A \cdot (c_0 + e_0) - (c_1 + e_1) =A \cdot (c_0 + e_0) - (A \cdot c_0 + e_1)  =A\cdot e_0 -e_1.$$ 

We would like to show that $z=e_0=w_0-c_0$. 
To this end, fix $i \in [k]$, we shall show that $z_i = (e_0)_i$. Let $S := \bigcup_{\ell \in \supp(e_0)\setminus \{i\}} S_\ell$. Note that for any $j \in S_i$ so that $j \notin S \cup \supp(e_1)$, we have that  $y_j = (e_0)_i$. We shall show that the majority of $j \in S_i$ satisfy that  $j \notin S \cup \supp(e_1)$, and consequently $z_i = (e_0)_i$. 
To see the latter, note that since $A$ is a $(d,b)$-design matrix, we have that $|S_i| \geq d$. Moreover, 
$|S_\ell \cap S_i| \leq  b$ for any $\ell \neq i$, and so 
$$|S \cap S_i| \leq \sum_{\ell \in \supp(e_0)\setminus \{i\}} |S_\ell \cap S_i|  \leq \wt(e_0) \cdot b.$$ But this implies in turn that 
$$|(S \cup \supp(e_1))\cap S_i| \leq |S\cap S_i| + |\supp(e_1)| \leq 
\wt(e_0) \cdot b+ \wt(e_1) \leq  \wt(e) \cdot b< \frac{d} {2}.$$
So the majority of $j \in S_i$ satisfy that  $j \notin S \cup \supp(e_1)$, and so $z_i = (e_0)_i$.

\begin{flushright} $\blacksquare$ \end{flushright}

Once more, by the definition of $D_k$ and Lemmas \ref{lem:sidon_2} and \ref{lem:sidon_3} we conclude that $D_k$ can be decoded from less than $\frac{d} {4}$ errors in time $\poly(k, \log q)$.

\section{Decoding double-circulant codes based on cyclic codes}\label{sec:cyc_to_circ}

In this section, we present an efficient decoding algorithm from square-root errors for the double-circulant codes based on cyclic codes from \cite[Theorem 1]{CPW69}.  Along the way, we also provide a more detailed proof for the distance  of these codes than the one that appears in \cite{CPW69}, and we also show that the transformation can be instantiated to obtain an asymptotic family of double-circulant codes of square-root distance.

\begin{theorem}[Efficient decoding of double-circulant codes based on cyclic codes]\label{thm:cyc_to_circ}
Let $q$ be a prime power, and let $I \subseteq \N$ be an infinite sequence of integers. Suppose that there exists an explicit family of codes 
$\mathcal{C}=\{C_k\}_{k\in I}$, where $C_k \subseteq \F_q^k$ is a cyclic code of distance $d=d(k)$ that can be decoded from $\frac d 2$ errors in time $T=T(k)$, 
and $C_k^\perp$ has distance $d^\perp=d^\perp(k)$ and can be decoded from $\frac {d^\perp} 2$ errors in time $T^\perp=T^\perp(k)$.
Then there exists an explicit family of codes $\mathcal{D}=\{D_k\}_{k\in I}$, where $D_k \subseteq \F_q^{2k}$ is a double-circulant code of distance at least $d':=\min\{d, d^\perp\}$ that can be decoded from $\frac {d'} 2$ errors in time $O(T+ T^\perp+\poly(k, \log q))$. 
Moreover,
 if $\mathcal{C}$ is $d$-balanced, then $\mathcal{D}$ is $(2,d')$-balanced. 
\end{theorem}

We prove the above theorem in Section \ref{subsec:cyc_to_circ} below. 
By instantiating this theorem with punctured RM codes, 
we deduce the following.

\begin{corollary}\label{cor:cyc_to_circ}
Let $I := \{2^m -1 \mid m \in \N\}$. 
Then there exists an explicit family of codes $\mathcal{D}=\{D_k\}_{k\in I}$, where $D_k \subseteq \F_2^{2k}$ is a double-circulant code of distance at least $d:=\Omega(\sqrt{k})$ that can be decoded from $\frac d 2$ errors  in time $\poly(k)$. 
Moreover, $\mathcal{D}$ is $(2,d)$-balanced.
 \end{corollary}

\proof 
For $k=2^m -1 \in I$, let $C_k \subseteq \F_2^k$ be the \emph{dual}  of the punctured RM code $\RM^*(\frac m 2, m)$.
 By   Fact \ref{fact:cyclic_dual} and Proposition \ref{prop:prm}, both $C_k = (\RM^*(\frac m 2, m))^\perp$ and $C_k^\perp = \RM^*(\frac m 2, m)$ are cyclic codes. 
By Fact \ref{fact:rm}, $C_k^\perp = \RM^*(\frac m 2, m)$ has distance at least $d^\perp:=2^{\frac m 2} -1= \Omega(\sqrt{k})$, and is decodable from $\frac {d^\perp} 2$-errors in time $\poly(k)$. Further observe that $C_k = (\RM^*(\frac m 2, m))^\perp$ is the code consisting of all codewords  in $\RM^*(\frac m 2 -1, m)$ which correspond to polynomials $f(x_1, \ldots, x_m)$ with $f(0)=0$. Therefore by Fact \ref{fact:rm},  $C_k$ has distance at least $d:=2^{m/2+1}-1 \geq \sqrt{k}$, and is decodable from $\frac d 2$ errors in time $\poly(k)$. 
 
Finally, we note that $C_k $ is $d$-balanced. To see this, let $c$ be a codeword in $C_k = (\RM^*(\frac m 2, m))^\perp$ associated with some non-zero polynomial $f(x_1, \ldots, x_m)$ of degree at most $\frac m 2-1$ which satisfies that $f(0)=0$. Then by the distance property, $c$ has weight at least $d$, and so $c$ has at least $d$ non-zero entries.  Furthermore, we have that $1-f(x_1, \ldots, x_m)$ is also a polynomial of degree at most $\frac m 2 -1$ which satisfies that $f(0)=1$, and so it is not the zero polynomial. So the codeword $\bar c \in \RM^*(\frac m 2 -1, m)$ associated with this polynomial has weight at least $d$. 
But this implies in turn that $c$ has at 
 least $d$ zero entries. So we have shown that any non-zero codeword in $C_k$ has at least $d$ non-zero entries, and at least $d$ zero entries, and consequently $C_k$ is $d$-balanced. 
 
The conclusion then follows by applying Theorem \ref{thm:cyc_to_circ} to the code family $\mathcal{C}$. 

\begin{flushright} $\blacksquare$ \end{flushright}

\begin{remark}
We could have also instantiated Theorem \ref{thm:cyc_to_circ} with BCH codes \cite{BR60}, and achieve double-circulant codes with parameters similar to those stated in Corollary \ref{cor:cyc_to_circ}. However, since dual BCH are not known to be efficiently decodable, this would not lead to efficient decoding algorithms for the resulting double-circulant codes. 
\end{remark}

\subsection{Proof of Theorem \ref{thm:cyc_to_circ}}\label{subsec:cyc_to_circ}

\paragraph{Construction.}
Fix $k \in I$, the code $D_k$ is constructed as follows.
Let $g(x) \in \F_q[x]$ be the generator polynomial for $C_k \subseteq \F_q^k$, and note that $\deg(g(x)) < k$. Let $a \in \F_q^k$ be the length $k$ coefficient vector of $g(x)$, viewed as a polynomial in $(\F_q)_{<k}[x]$, and let $A$ be the $k \times k$ circulant matrix whose first column is $a$. Finally, let $D_k$ be a double-circulant code with generator matrix $G=\begin{pmatrix}
        I_k \\ A
    \end{pmatrix}$.

\paragraph{Distance and balanced weight.}

Let $c=G\cdot m$ for a non-zero $m \in \F_q^k$, we shall show that $\wt(c) \geq \min \{d,d^\perp\}$.
Let $h(x) \in \F_q[x]$ be the check polynomial for $C_k$.
Let us write $m(x)$ as $m(x)=f(x)\cdot h(x)+r(x)$, where $f(x), r(x) \in \F_q[x]$ and $\deg(r(x))<\deg(h(x))$. Next we divide into two cases. 

\paragraph{Case 1 -- $r(x)\neq 0$:}

We shall show that in this case $\wt(A \cdot m) \geq d$, and so $\wt(c) = \wt(m, A\cdot m) \geq d$. 

By Claim \ref{clm:circulant} and the structure of $A$, we have that $A \cdot m$ is the coefficient vector of $g(x)\cdot m(x) \; (\bmod\; x^k -1)$. Furthermore, since $g(x)\cdot h(x)=x^k-1$, we have that
$$
g(x)\cdot m(x) \equiv g(x)\cdot(f(x)\cdot h(x)+r(x)) \equiv g(x)\cdot r(x) \; (\bmod\; x^k -1),
$$
where 
$$\deg(g(x) \cdot r(x))=\deg(g(x))+\deg(r(x)) < \deg(g(x))+ \deg(h(x)) = \deg(g(x)\cdot h(x)) = k.$$ So we conclude that $A \cdot m$ is the length $k$ coefficient vector of $g(x) \cdot r(x)$. But by assumption that $r(x) \neq 0$,  we have that the coefficient vector of $g(x) \cdot r(x)$ is a non-zero codeword of $C_k$, and so has weight at least $d$, which implies in turn that $\wt(A \cdot m) \geq d$. 

\paragraph{Case 2 -- $r(x) = 0$:}

In this case, we shall show that $\wt(m) \geq d^\perp$, and so $\wt(c) = \wt(m, A\cdot m) \geq d^\perp$.

By assumption that $r(x)=0$, we have that $m(x) = f(x) \cdot h(x)$, and so $m \in \F_q^k$ is a non-zero codeword in the cyclic code whose generator polynomial is $h(x)$. Let $k' := k-\deg(g)$. By Fact \ref{fact:cyclic_dual}, we have that $h^\rev(x) \in (\F_q)_{\leq k'}[x]$ generates $C^\perp$. Lemma \ref{Dual_Check} below then implies that $h(x)$ 
 generates the code $(C^\perp)^\rev$. So we conclude that $m(x)$ is a non-zero codeword in  $(C^\perp)^\rev$, and consequently $\wt(m) \geq d^\perp$. 
    
\begin{lemma}\label{Dual_Check}
Let $C \subseteq \F^n$ be a cyclic code of dimension $k$ with generator polynomial $g(x) \in \F_{\leq n-k}[x]$. Then $g^\rev(x) \in \F_{\leq n-k}[x]$ is a generator polynomial for $C^\rev$. 
\end{lemma}

\proof
It suffices to show that for any polynomial $f(x) \in \F_{<k}[x]$, 
$$
  (g(x)\cdot f(x))^\rev =  g^\rev(x)\cdot f^\rev(x),
$$
where we view $g(x)\cdot f(x)$ as a polynomial in $\F_{<n}[x]$. 

But the above follows since for any $f(x) \in \F_{<k}[x]$,
$$(g(x)\cdot f(x))^\rev  = x^{n-1} \cdot g\left(\frac 1 x\right) \cdot f\left(\frac 1 x\right) = \left(x^{n-k} \cdot g\left(\frac 1 x\right)  \right) \cdot \left(x^{k-1} \cdot f\left(\frac 1 x\right)  \right) = g^\rev(x)\cdot f^\rev(x).$$

\begin{flushright} $\blacksquare$ \end{flushright}

Finally, note that if  $C_k$ is $d$-balanced, then in the case that $r(x) \neq 0$, we have that $A \cdot m$ is a  non-zero codeword of $C_k$, and so $\wt_\bal(A \cdot m) \geq d$, while in the case that $r(x)=0$, we have that $\wt(m) \geq d^\perp$. So we have that $\wt(m)+\wt_\bal(m) \geq \min\{d, d^\perp\}$, and hence $D_k$ is $(2,d')$-balanced.

\paragraph{Decoding.}

Let $\Dec$, $\Dec^\perp$ denote the decoding algorithms for $C_k,C_k^\perp$, respectively. The decoding algorithm $\Dec'$ for $D_k$ is presented in Figure \ref{fig:cyclic_dec} below. In what follows,  let $g(x) \in (\F_q)_{<k'}(x), h(x) \in (\F_q)_{<k-k'}(x)$ denote the generator and check polynomial for $C_k$, respectively, where $k' = \dim(C_k)$,

\begin{figure}[h]
  \begin{boxedminipage}{\textwidth} \small \medskip \noindent
    $\;$

    \underline{\textbf{Decoding algorithm $\Dec'$  for the double-circulant code $D_k$:}}
\begin{itemize}
\item \textbf{INPUT:} A received word $w=(w_0, w_1) \in \F_q^{2k}$, where $w_0, w_1 \in \F_q^k$ 
\item \textbf{OUTPUT:}  A codeword $c \in C_k$ so that $\dist(w,c)< \frac{d'} {2}$, or $\bot$ if such a codeword does not exist

\end{itemize}

\begin{enumerate}
    \item \label{step:cyclic_codeword1} Let $c_1 = \Dec(w_1) \in \F_q^k$, and let $r(x) = \frac{c_1(x)} {g(x)} \in (\F_q)_{\leq k}[x]$.  ; If $\Dec$ outputs $\bot$, then return $\bot$. 
    \item \label{step:cyclic_codeword2} Let 
    $c_0 = (\Dec^\perp( (w_0-r)^{\rev} ))^{\rev} \in \F_q^k$; If $\Dec^\perp$ outputs $\bot$, then return $\bot$. 
    \item \label{step:cyclic_output} Let $c = G\cdot (c_0+r)$.  
    If $\dist(c, w)< \frac {d'} {2}$, return $c$. 
\end{enumerate}

  \medskip

  \end{boxedminipage}

\caption{Decoding algorithm for the double-circulant code $\mathcal{D}$}
\label{fig:cyclic_dec}
\end{figure}

The running time of the algorithm is clearly as stated. 
To show correctness, let $w=(w_0, w_1) \in \F_q^{2k}$ be a received word, where $w_0, w_1 \in \F_q^k$, and let 
$c=(m, A\cdot m) \in C$ be a codeword so that $\dist(c,w) < \frac{d'} 2$. We shall show that the algorithm will output $c$.

As before, we shall write $m(x)$ as $m(x)=f(x)\cdot h(x)+r'(x)$, where $f(x), r'(x) \in \F_q[x]$ and $\deg(r'(x))<\deg(h(x))$. By the analysis of the distance above we have that $A \cdot m$ is the coefficient vector of $g(x) \cdot r'(x)$. In particular, this means that $A \cdot m$ is a codeword of $C_k$ so that  
$\dist(A \cdot m, w_1) < \frac{ d'} 2 \leq \frac d 2$, and so on Step \ref{step:cyclic_codeword1}, we shall have $c_1 = A \cdot m$, $c_1(x) = g(x) \cdot r'(x)$, and $r(x)= r'(x)$.

Next observe that  
$$ (m(x) - r(x))^\rev = (f(x) \cdot h(x))^\rev = h^\rev(x) \cdot f^\rev(x),$$
where we view $f(x) \cdot h(x)$ as a polynomial in $(\F_q)_{<k}[x]$.
In particular, this means that $(m-r)^\rev$ is a codeword of $C_k^\perp$ so that 
$$\dist( (m-r)^\rev, (w_0 -r)^\rev) = \dist(m,w_0)< \frac{d'} {2} \leq \frac{d^\perp} 2,$$ and so on Step \ref{step:cyclic_codeword2}, we shall have $c_0 = m-r$.

Finally, the above implies in turn that  $c_0+r=m$, and consequently on Step \ref{step:cyclic_output}, the algorithm shall output the codeword $c$.

\section{Decoding Wozencraft Codes}\label{sec:circ_to_wozen}

In this section, we first describe in Section \ref{subsec:circ_to_wozen} below a transformation from double-circulant to Wozencraft codes that is implicit in \cite{GL25}, and show that this transformation also preserves efficiency of decoding. Then in Section \ref{subsec:sidon_to_wozen}, we show how to instantiate this transformation with the results of Section \ref{sec:Sidon_to_DC} to obtain 
an explicit construction of a Wozencraft code that can be efficiently decoded from square-root errors. Finally, in Section \ref{subsec:cyc_to_circ_limit}, we discuss limitations on instantiating this transformation with our results of Section \ref{sec:cyc_to_circ}, as well as with the Fourier-based double-circulant codes of \cite{RL90}.

\subsection{Decoding Wozencraft codes based on double-circulant codes}\label{subsec:circ_to_wozen}

In this section, we present a transformation that turns double-circulant codes  of certain block lengths into Wozencraft codes with roughly the same distance. 
This transformation was implicit in \cite{GL25}. We make it explicit, and further show that this transformation preserves efficiency of decoding. For the sake of completeness, we present a more general transformation which turns $t$-circulant codes into $t$-Weldon codes for any $t \geq 2$.

\begin{lemma}[From circulant codes to Weldon Ensemble]\label{Tcirc_to_TWozen}
Let $q$ be a prime power, and let $I \subseteq \N$ be an infinite sequence of primes so that $q$ is a primitive root modulo any $k \in I$.
Suppose that there exists an explicit family of codes $\mathcal{D}=\{D_k\}_{k\in I}$, where $D_k \subseteq \F_q^{t \cdot k}$ is a $(t,d)$-balanced $t$-circulant code for $d = d(k)$. 
Then there exists an explicit family of codes $\mathcal{W}=\{W_k\}_{k \in I}$, where $W_k \subseteq \F_q^{t \cdot (k-1)}$ is a $t$-Weldon code of distance  at least $d$. 
Moreover, if $\mathcal{D}$ can be decoded from $\frac d 2$ errors in time $T=T(k)$,  then $\mathcal{W}$ can be decoded from $\frac d 2$ errors in time $O(q^{t-1}\cdot (T+\poly(t,k, \log q))$.
\end{lemma}

Next we describe the construction of the Weldon Ensemble from circulant codes, and then analyze its distance and present a decoding algorithm for it. 

\paragraph{Construction.}

Fix $k \in I$, the code $W_k$ is constructed as follows. 

    By Proposition \ref{prop:primitive}, and by assumption that $k$ is a prime and that $q$ is a primitive root modulo $k$, we have that $p_k(x) := \sum_{i=0}^{k-1}x^i$ is an irreducible polynomial over $\F_q$.
    Therefore we have that $\bH:=\F_q[x] \big/ p_k(x)$ is a finite field of $q^{k-1}$ elements that is isomorphic to $\F_{q^{k-1}}$. Further note that the map $\varphi: \bH \to \F_q^{k-1}$ which maps a polynomial $f(x) \in \bH$  to its length $k-1$ coefficient vector is an $\F_q$-linear bijection. It thus suffices to find  $\alpha_1(x),\ldots,\alpha_{t-1}(x) \in \bH$ so that 
    $$W_k=\bigg\{ \big(\varphi(m(x)), \varphi(\alpha_1(x) \cdot m(x)), \ldots, \varphi(\alpha_{t-1}(x) \cdot m(x))\big) \mid m(x) \in \bH \bigg\},$$
    where multiplication is performed over the field $\bH$. 

   The polynomials $\alpha_i(x)$ are defined as follows. 
    Since $D_k \subseteq \F_q^{t \cdot k}$ is a $t$-circulant code, there exist $k \times k$ circulant matrices $A_1, \ldots, A_{t-1}$ over $\F_q$ so that $$G=\begin{pmatrix}
        I_k \\ A_1 \\ \vdots \\ A_{t-1}
    \end{pmatrix}$$
    is a generator matrix for $D_k$.  For $i \in [t-1]$, let $a^{(i)} \in \F_q^k$ be the first column of $A_i$, and  let $\alpha_i(x):= a^{(i)}(x) \; (\bmod \; p_k(x)) \in \bH$. 

\paragraph{Distance.}

Let $c=(c_0, c_1, \ldots, c_{t-1}) \in \F_q^{t \cdot (k-1)}$ be a non-zero codeword of $W_k$ which corresponds to some non-zero $m(x) \in \bH$, where $c_i \in \F_q^{k-1}$ for any $i \in \{0,1,\ldots, t-1\}$. We shall show that $\wt(c) \geq d$. 

For $i \in [t-1]$, let 
$$f_i(x):=a^{(i)}(x) \cdot m(x) \; (\bmod \; p_k(x)) \in (\F_q)_{<k-1}[x],$$
and let $$g_i(x):=a^{(i)}(x) \cdot m(x) \; (\bmod \; x^k-1) \in (\F_q)_{<k}[x].$$

Note that by the definition of  $W_k$, $c_0$ is the length $k-1$ coefficient vector of $m(x) \in \bH$, and for any $i \in [t-1]$,  $c_i$ is the length $k-1$ coefficient vector of $f_i(x)$.
On the other hand, if we let $c'_0 \in \F_q^k$ be the length $k$ coefficient vector of $m(x)$ (viewed as a polynomial in $(\F_q)_{<k}(x)$), and $c'_i \in \F_q^k$ be the length $k$ coefficient vector of $g_i(x)$ for $i \in [t-1]$, then  by Claim \ref{clm:circulant},
$c':=(c'_0, c'_1, \ldots, c'_{t-1}) \in \F_q^{t \cdot k}$ is a non-zero codeword of the $t$-circulant code $D_k$. We clearly have that $\wt(c_0)=\wt(c'_0)$. We shall show that $\wt(c_i) \geq \wt_\bal(c'_i)$ for any $i \in [t-1]$, and so $\wt(c) \geq \wt(c'_0) + \sum_{i=1}^{t-1}\wt_\bal(c'_i)\geq d$, where the right-hand inequality follows by our assumption that $D_k$ is $(t,d)$-balanced. 

Fix $i \in [t-1]$, we shall show that $\wt(c_i) \geq \wt_\bal(c'_i)$.
To see this, note that since $p_k(x)$ divides $x^k -1$, we have that $f_i(x) = g_i(x) \; (\bmod \; p_k(x))$. 
Therefore, by the structure of $p_k(x)$,  $c_i$ can be obtained from $c'_i$ by removing the last entry of $c'_i$ and subtracting it from all other entries of $c'_i$.  But this means that if we denote by $\beta \in \F_q$ the last entry of $c_i$, then the number of zeros in $c_i$ equals the number of occurrences of $\beta$ in $c'_i$ minus $1$, which is at most $k-\wt_\bal(c'_i)-1$. Since $c_i$ has length $k-1$, this implies in turn that
$\wt(c_i) \geq \wt_\bal(c'_i).$

\paragraph{Decoding.} 
Assume we have a decoding algorithm $\Dec$ for the $t$-circulant code $\mathcal{D}$ from $\frac d 2$ errors. 
The decoding algorithm $\Dec'$ for the $t$-Weldon code $\mathcal{W}$ is given in Figure \ref{fig:weldon_dec} below.

\begin{figure}[h]
  \begin{boxedminipage}{\textwidth} \small \medskip \noindent
    $\;$

    \underline{\textbf{Decoding algorithm  $\Dec'$ for the $t$-Weldon Code $W_k$:}}
\begin{itemize}
\item \textbf{INPUT:} A received word $w=(w_0, w_1, \ldots, w_{t-1}) \in \F_q^{t \cdot (k-1)}$, where $w_i \in \F_q^{k-1}$ for any $i \in \{0,1,\ldots, t-1\}$
\item \textbf{OUTPUT:}  A codeword $c \in W_k$ so that $\dist(w,c)< \frac{d} {2}$, or $\bot$ if such a codeword does not exist

\end{itemize}

    \begin{enumerate}
\item For any $\beta_1, \ldots, \beta_{t-1} \in \F_q$:
\begin{enumerate}
\item \label{step:weldon_dec_w'} Let $w':=(w'_0, w'_1, \ldots, w'_{t-1}) \in \F_q^{t \cdot k}$, where $w'_0 = w_0 \circ 0 \in \F_q^k$, and $w'_i = w_i \circ 0 + \beta_i \cdot 1_k \in \F_q^k$ for any $i \in [t-1]$, and where $1_k$ denotes the length $k$ all ones string.
\item \label{step:weldon_dec_A'} Apply the decoding algorithm $\Dec$ for $D_k$ on $w'$. If $\Dec$ outputs $\bot$, then continue to the next iteration; Otherwise, let $c' \in \F_q^{t \cdot k}$ be the output of $\Dec$.
\item
\label{step:weldon_dec_c}
Let $c := (c_0, c_1, \ldots, c_{t-1}) \in \F_q^{t \cdot (k-1)}$, where $c_0 \in \F_q^{k-1}$ is obtained from $c'_0$ by removing the last entry of $c'_0$, and for any $i \in [t-1]$, $c_i \in \F_q^{k-1}$ is obtained from $c'_i$ by removing the last entry of $c'_i$ and subtracting it from the rest of the entries of $c'_i$. 
\item  \label{step:weldon_dec_output} If $c \in W_k$ and $\dist(c, w) < \frac d 2$, then return $c$ and halt. 
\end{enumerate}
\item Return $\bot$.
\end{enumerate}

  \medskip

  \end{boxedminipage}

\caption{Decoding algorithm for the $t$-Weldon Code $\mathcal{W}$}
\label{fig:weldon_dec}
\end{figure}

The running time of the algorithm is clearly as stated. 
To show correctness, assume that $ c = ( c_0,   c_1, \ldots,  c_{t-1}) \in W_k$ is a codeword so that $\dist( c,w) < \frac d 2$, where $ c_i \in \F_q^{k-1}$ for any $i \in \{0,1,\ldots, t-1\}$. It suffices to show that there exist $\beta_1, \ldots, \beta_{t-1} \in \F_q$ so that on the iteration corresponding to $\beta_1, \ldots, \beta_{t-1}$, the algorithm $\Dec'$ will output $ c$. 
To see this, recall that by the Definition of  $W_k$, there exists $m(x) \in (\F_q)_{<k-1}[x]$ so that $ c_0$ is the length $k-1$ coefficient vector of $m(x)$, and $ c_i$ is the length $k-1$ coefficient vector of  
$f_i(x):=a^{(i)}(x) \cdot m(x)\;(\bmod \; p_k(x))$ for any $i \in [t-1]$. For $i \in [t-1]$, let $\beta_i \in \F_q$ be the coefficient of $x^{k-1}$ in $g_i(x):=a^{(i)}(x) \cdot m(x)\;(\bmod \; x^k -1)$. We shall show that $ c$ will be output in the iteration corresponding to $\beta_1, \ldots, \beta_{t-1}$. 

To see the latter, let $c'=(c'_0, c'_1, \ldots, c'_{t-1}) \in \F^{t \cdot k}$, where $c'_0 \in \F_q^k$ is the length $k$ coefficient vector of $m(x)$ (viewed as a polynomial in $(\F_q)_{<k}(x)$), and 
$c'_i \in \F_q^k$ is the length $k$ coefficient vector of $g_i(x)$ for $i \in [t-1]$. Then we have that $c' \in D_k$ and $c'_0 = c_0 \circ 0$. Moreover, since $p_k(x)$ divides $x^k -1$, for any $i \in [t-1]$, we have that
$f_i(x) = g_i(x)\;(\bmod \; p_k(x))$, and consequently, $c'_i  =  c_i \circ 0 + \beta_i \cdot 1_k$. But this implies in turn that $\dist(c',w') < \frac d 2$, and so the algorithm $\Dec$ will output $c'$ on Step \ref{step:weldon_dec_A'}. Finally, by the definition of $c'$, this also implies that the algorithm $\Dec'$ will compute $ c$ on Step \ref{step:weldon_dec_c}.

\subsection{Efficiently decodable Wozencraft code}\label{subsec:sidon_to_wozen}

 Instantiating Lemma \ref{Tcirc_to_TWozen} with the double-circulant codes based on Sidon sets given by Corollary \ref{cor:DC} implies the following corollary. 
    
\begin{corollary}
Let $q$ be a prime power, and let $I \subseteq \N$ be an infinite sequence of primes so that $q$ is a primitive root modulo any $k \in I$.
Then there exists an explicit family of codes $\mathcal{W}=\{W_k\}_{k \in I}$, where $W_k \subseteq \F_q^{2 \cdot (k-1)}$ is a Wozencraft code of distance  at least $d=\Omega(\sqrt{k})$ that can be decoded from $\frac d 2$ errors in time $\poly(k,\log q)$.
\end{corollary}

A widely accepted conjecture by Artin states the following.
\begin{conjecture}[Artin’s Conjecture on Primitive Roots, \cite{HB86}]
    For any integer m that is neither a square number nor $-1$, it is a primitive root modulo infinitely many primes $p$. Moreover, the set of prime numbers $p$ such that $m$ is a primitive root modulo $p$ has a positive asymptotic density inside the set of all primes.
\end{conjecture}

    Assuming Artin's conjecture, for any prime $q$,  primes $k$ for which $q$ is a primitive root modulo $k$ exist infinitely often at sufficiently high density and can be efficiently found in deterministic $\poly(k)$ time.
    In  \cite{HB86}, Heath proved that a weaker version of Artin's conjecture holds for every prime $m$ except at most two primes. This shows that for all but at most two choices of the prime alphabet size $q$, primes $k$  such that $q$ is a primitive root modulo $k$ exist  infinitely often at sufficiently high density and can be efficiently found in deterministic $\poly(k)$ time.

\subsection{Limitations}\label{subsec:cyc_to_circ_limit}

In this section, we discuss several limitations on instantiating the transformation given by Lemma \ref{Tcirc_to_TWozen}  with the double-circulant codes based on cyclic codes of \cite{CPW69} or the Fourier-based double-circulant codes of \cite{RL90}. 

We begin by mentioning several limitations that follow by the requirement in Lemma \ref{Tcirc_to_TWozen} that $q$ is a primitive root modulu $k$. 
First, we note that the Fourier-based double-circulant codes of square-root distance, described in \cite[Example 1]{RL90}, satisfy that $k=q^s-1$ for some integer $s$, and therefore $q$ cannot be a primitive root modulo $k$. 
Next we note that for cyclic codes, the requirement that $q$ is a primitive root modulo $k$ implies that $\min\{d, d^\perp\} \leq 2$, and consequently Theorem \ref{thm:cyc_to_circ} cannot give double-circulant codes with non-trivial distance satisfying this requirement.

\begin{proposition}
Let $q$ be a prime power, and let $n$ be a prime so that $q$ is a primitive root modulo $n$. 
Then if $C \subseteq \F_q^n$ is a cyclic code, then $\min\{\Delta(C), \Delta(C^\perp)\}\leq 2$. 
\end{proposition}

\proof
By Proposition \ref{prop:primitive}, the assumption that $q$ is a primitive root modulo $n$ implies that 
$p_n(x)=\sum_{i=0}^{n-1}x^i$ is an irreducible polynomial over $\F_q$.
Consequently, 
 $x^n-1=(x-1)\cdot p_n(x)$ is the only non-trivial way to decompose $x^n-1$. 
 
 Let $g(x), h(x)$ denote the generator and check polynomial of $C$, respectively. Then $g(x) \cdot h(x) = x^n-1$, and consequently either $g(x)$ or $h(x)$ divides $x-1$. In the case that $g(x)$ divides $x-1$, we have that either $g(x)=1$, in which case $C$ is the identity code of distance $1$, or that $g(x)=x-1$, in which case $C$ is the parity code of distance $2$. In the case that $h(x)$ divides $x-1$, we have that either $h^\rev(x) =1$, in which case $C^\perp$ is the identity code of distance $1$, or that $h^\rev(x) = -x+1$, in which case $C^\perp$ is the parity code of distance $2$. So we have that in either cases, $\min\{\Delta(C), \Delta(C^\perp)\}\leq 2$.
\begin{flushright} $\blacksquare$ \end{flushright}

Consequently, we cannot instantiate the transformation given by  Lemma \ref{Tcirc_to_TWozen} with either the Fourier-based double-circulant codes of \cite{RL90} or the double-circulant codes based on cyclic codes of \cite{CPW69}.
One possible approach to overcome these limitations may be to  replace in the proof of Lemma \ref{Tcirc_to_TWozen} the polynomial $p_k(x)=\sum_{i=0}^{k-1} x^i$ with another irreducible polynomial $r(x)$ of degree $s$, and let
 $$W_k=\bigg\{ \big(\varphi(m(x)), \varphi(\alpha(x) \cdot m(x))) \mid m(x) \in \bH \bigg\},$$
 where $\bH:=\F_q[x] \big/ r(x)$, $\varphi: \bH \to \F_q^s$  is the $\F_q$-linear bijection which maps a polynomial $f(x) \in \bH$  to its length $s$ coefficient vector, and $\alpha(x)$ is defined as in the proof of Lemma \ref{Tcirc_to_TWozen}. 
 The proof of Lemma \ref{Tcirc_to_TWozen} then requires that $r(x)$ divides $x^k-1$, since it relies on the fact that for a polynomial $f(x)$, $f(x) \; (\bmod \;r(x)) = (f(x) \; (\bmod \;x^k-1) )\; (\bmod \;r(x))$.
 The next proposition shows that for certain values of $k$, the latter requirement implies that $r(x)$ is of low degree. 

\begin{proposition}
Let $q$ be a prime power, and let $k,s$ be integers so that $q^s \equiv 1 \; (\bmod \; k)$. Then if $r(x)$ is an irreducible polynomial over $\F_q$ which divides $x^k-1$, then $\deg(r(x))\leq s$. 
\end{proposition}

\proof
By assumption that $q^s \equiv 1 \; (\bmod \; k)$, we have that $k \mid q^s -1$, and so $x^k -1 \mid x^{q^s-1} -1$. By assumption that $r(x) \mid x^k-1$, this implies in turn that $r(x) \mid x^{q^s-1} -1$. Next recall that the roots of $x^{q^s-1} -1$ are all elements in the multiplicative group of the field $\F_{q^s}$. Since $r(x) \mid x^{q^s-1} -1$, the same holds for all roots of $r(x)$, and since $r(x)$ is irreducible over $\F_q$, this implies in turn that $\deg(r(x))\leq s$. 
\begin{flushright} $\blacksquare$ \end{flushright}

We note that the Fourier-based double-circulant codes of \cite{RL90} must satisfy the requirement that $q^s \equiv 1 \; (\bmod \; k)$ for some integer $s$. Moreover, as mentioned above, the Fourier-based double-circulant codes of square-root distance given in \cite[Example 1]{RL90} satisfy that $k=q^s-1$ for some integer $s$, and so by the above proposition, $\deg(r(x)) \leq s\leq \log(k)$, in which case the Wozencraft code obtained using the polynomial $r(x)$ 
can have distance at most $\log(k)$. 
Likewise, the double-circulant codes obtained by instantiating Theorem \ref{thm:cyc_to_circ} with either punctured RM or BCH codes must also satisfy the requirement that  $q^s \equiv 1 \; (\bmod \; k)$ for some integer $s$, and in the special case in which the punctured RM or BCH codes are evaluated over all non-zero elements in the field then $k=q^s-1$ for some integer $s$. Consequently, the same limitations as before apply. 

\medskip

Finally, we note that even without the restriction on the block length, there may be limitations on the distance of the double-circulant and Wozencraft codes based on cyclic codes. 
Specifically, we recall the following conjecture on the non-existence of asymptotically good cyclic codes.

\begin{conjecture}\label{conj:cyclic}
For any fixed prime power $q$, there does not exist an infinite family of asymptotically good cyclic codes over $\F_q$. 
\end{conjecture}

We show that assuming the above conjecture, there do not exist cyclic codes with $\min\{d, d^\perp\}= \Omega(n)$.

\begin{proposition}\label{prop:cyclic_conj}
Assuming Conjecture \ref{conj:cyclic}, for any fixed prime power $q$, and for any infinite family $\mathcal{C}=\{C_n\}_n$, where $C_n \subseteq \F_q^n$ is a cyclic code, we have that $\min\{\Delta(\mathcal{C}), \Delta(\mathcal{C}^\perp) \} = o(n)$. 
\end{proposition}

\proof 
Let us assume by contradiction that $\min\{\Delta(\mathcal{C}), \Delta(\mathcal{C}^\perp) \} = \Omega(n)$. We know that for any $n \in \N$, $R(C_n)+R(C_n^\perp)=1$, therefore either $R(C_n)\geq 1/2$ or $R(C_n^\perp)\geq 1/2$. But since both $C_n$ and $C_n^\perp$ are cyclic codes, 
this implies in turn the existence of an infinite family of cyclic codes of rate $ \frac 1 2$ and constant relative distance, contradicting Conjecture \ref{conj:cyclic}. 
\begin{flushright} $\blacksquare$ \end{flushright}

In particular, assuming Conjecture \ref{conj:cyclic}, the transformations  given by Theorem \ref{thm:cyc_to_circ} cannot lead to double-circulant of distance $\Omega(k)$.

  \paragraph{Acknowledgement.} We thank Venkatesan Guruswami and Mary Wootters for helpful discussions.

\bibliographystyle{alpha}
\bibliography{bib}

\begin{thebibliography}{CPW69}

\bibitem[BC62]{BC60}
R.~C. Bose and Sarvadaman Chowla.
\newblock Theorems in the additive theory of numbers.
\newblock {\em Commentarii Mathematici Helvetici}, 37:141--147, 1962.

\bibitem[BR60]{BR60}
R.~C. Bose and Dwijendra~K. Ray{-}Chaudhuri.
\newblock On {A} class of error correcting binary group codes.
\newblock {\em Inf. Control.}, 3(1):68--79, 1960.

\bibitem[BTS74]{BTS74}
Vijay Bhargava, Stafford Tavares, and Saligram Shiva.
\newblock Difference sets of the hadamard type and quasi-cyclic codes.
\newblock {\em Information and Control}, 26(4):341--350, 1974.

\bibitem[Cal83]{Calder83}
Robert Calderbank.
\newblock A square root bound on the minimum weight in quasi-cyclic codes.
\newblock {\em {IEEE} Transactions on Information Theory}, 29(3):332--336,
  1983.

\bibitem[Ceb50]{Cheb1850}
Pafnutij~Lvoviˇc Cebyˇsev.
\newblock M´emoire sur les nombres premiers, 1850.

\bibitem[CPW69]{CPW69}
C.~L. Chen, W.~Wesley Peterson, and E.~J. Weldon.
\newblock Some results on quasi-cyclic codes.
\newblock {\em Information and Control}, 15(5):407--423, 1969.

\bibitem[GL25]{GL25}
Venkatesan Guruswami and Shilun Li.
\newblock A deterministic construction of a large distance code from the
  wozencraft ensemble.
\newblock {\em {IEEE} Trans. Inf. Theory}, 71(2):930--935, 2025.

\bibitem[HB86]{HB86}
Roger Heath-Brown.
\newblock Artin's conjecture for primitive roots.
\newblock {\em The Quarterly Journal of Mathematics}, 37(1):27--38, 03 1986.

\bibitem[Jus72]{Just72}
J{\o}rn Justesen.
\newblock Class of constructive asymptotically good algebraic codes.
\newblock {\em {IEEE} Trans. Inf. Theory}, 18(5):652--656, 1972.

\bibitem[Kas74]{Kasami74}
Tadao Kasami.
\newblock A gilbert-varshamov bound for quasi-cycle codes of rate 1/2
  (corresp.).
\newblock {\em {IEEE} Transactions on Information Theory}, 20(5):679, 1974.

\bibitem[KLP68]{KLP68}
Tadao Kasami, Shu Lin, and W.~Wesley Peterson.
\newblock New generalizations of the reed-muller codes-i: Primitive codes.
\newblock {\em {IEEE} Trans. Inf. Theory}, 14(2):189--199, 1968.

\bibitem[Mas63]{Massey63}
James~L Massey.
\newblock Threshold decoding, 1963.

\bibitem[MS77]{MS77}
Jessie MacWilliams and Neil Sloane.
\newblock {\em The Theory of Error-Correcting Codes}.
\newblock North-holland Publishing Company, 1977.

\bibitem[Mul54]{Muller54}
David~E. Muller.
\newblock Application of boolean algebra to switching circuit design and to
  error detection.
\newblock {\em Transactions of the I.R.E. Professional Group on Electronic
  Computers}, 3(3):6--12, 1954.

\bibitem[PZ20]{PZ20}
Aditya Potukuchi and Yihan Zhang.
\newblock Improved efficiency for covering codes matching the sphere-covering
  bound.
\newblock In {\em {IEEE} International Symposium on Information Theory, {ISIT}
  2020, Los Angeles, CA, USA, June 21-26, 2020}, pages 102--107. {IEEE}, 2020.

\bibitem[Ree54]{Reed54}
Irving~S. Reed.
\newblock A class of multiple-error-correcting codes and the decoding scheme.
\newblock {\em Transactions of the I.R.E. Professional Group on Information
  Theory}, 4:38--49, 1954.

\bibitem[RL90]{RL90}
Ron~M. Roth and Abraham Lempel.
\newblock Application of circulant matrices to the construction and decoding of
  linear codes.
\newblock {\em {IEEE} Transactions on Information Theory}, 36(5):1157--1163,
  1990.

\bibitem[Shp13]{Shpilka13}
Amir Shpilka.
\newblock New constructions of {WOM} codes using the wozencraft ensemble.
\newblock {\em {IEEE} Trans. Inf. Theory}, 59(7):4520--4529, 2013.

\bibitem[Shp14]{Shpilka14}
Amir Shpilka.
\newblock Capacity-achieving multiwrite {WOM} codes.
\newblock {\em {IEEE} Trans. Inf. Theory}, 60(3):1481--1487, 2014.

\bibitem[Sid32]{Sidon32}
Simon Sidon.
\newblock Ein satz über trigonometrische polynome und seine anwendung in der
  theorie der fourier-reihen.
\newblock {\em Mathematische Annalen}, 106:DXXXVI--DXXXIX, 1932.

\bibitem[Wel73]{Weldon73}
E.~J. Weldon.
\newblock Justesen's construction-the low-rate case (corresp.).
\newblock {\em {IEEE} Trans. Inf. Theory}, 19(5):711--713, 1973.

\end{thebibliography}

\end{document}